# The CUFF, Clenching Upper-limb Force Feedback wearable device: design, characterization and validation.

Barontini, F.[3,*], Catalano, M.G.[3], Fani, S.[1,2], Grioli, G.[3], Bianchi, M.[1,2], Bicchi, A.[1,2,3]

[1] Department of Information Engineering, University of Pisa, Pisa, Italy

[2] Research Center "E. Piaggio", University of Pisa, Pisa, Italy

[3] Department of Soft Robotics for Human Cooperation and Rehabilitation, Istituto Italiano di Tecnologia, Genoa, Italy

*Corresponding author

E-mail: federicabarontini16@gmail.com

# Abstract

This paper presents the design, characterization and validation of a wearable haptic device able to convey skin stretch, force feedback, and a combination of both, to the user's arm. In this work, we carried out physical and perceptual characterization with eleven able-bodied participants as well as two experiments of discrimination and manipulation task hiring a total of 32 participants. In both the experiments the CUFF was used in conjunction with the Pisa/IIT SoftHand. The first experiment was a discrimination task where the subjects had to recognize the dimension and the softness between pair of cylinder. in the second experiment the subjects were asked to control the robotic hand for grasping objects. After the experiments the subjects underwent to a subjective evaluation of the device. Results of the experiments and questionnaire showed the effectiveness of the proposed device. Thank to its versatility and structure, the device could be a viable solution for teleoperation application, guidance and rehabilitation tasks, including prosthesis applications.

# Section I: Introduction

The sense of touch represents an astonishingly powerful information generator, where different types of data from the external world are continuously processed by the receptors distributed over human body [1].

Wearable haptic devices have shown promising results in the fields of robotic teleoperation [2], [3], rehabilitation [4], and guidance [5]. They can be easily and comfortably worn by the human user, be carried around and be integrated into their everyday life [6].

Furthermore, the possibility to integrate haptic system into the human body with minimal or no constraints to its motion [7], can be exploited to study the human behavior in a more ecological way. Typically, these devices can be applied to different parts of the human body, such as arms and forearms [7], feet [8] and fingers [9] and are mainly thought to deliver tangential and normal force, as well as texture and softness cues, see e.g. [10], [11]. Even though the fingertip is generally intended as the privileged channel for discriminative touch (see [11] for a review of wearable haptic system on a fingertip), forearm and arm are often chosen as an effective site for haptic feedback delivery. This choice is mandatory in those cases when the hand location is not physically available, e.g. for amputees, or when constraints on feedback location impose a different placement of the device, e.g. for haptic guidance of people with visual impairments, who are usually required to have their hands free to properly interact and to understand the external environment [12].

For these reasons, in literature, it is possible to find dif- ferent systems that deliver force information to the arm and forearm, especially tangential force for haptic guidance and proprioception. In [13] authors developed a wearable outer- covering haptic display (wOCHD) for hand motion, with two ball effectors to deform the skin and provide guiding information in four axes of motion. In [14] authors presented a wearable haptic feedback device for rotational skin stretch to the hairy skin, which results as an effective means of providing information about a user's controlled joint or limb motions for motion training and similar applications. A similar solution is presented in [15], and in [16] where a rotational skin stretch device is used to provide kinesthetic feedback from a prosthetic device. In [17] authors presented a lightweight bracelet with four servo motors to provide independent skin stretches at the palmar, dorsal, ulnar, and radial sides of the arm for haptic navigation tasks.

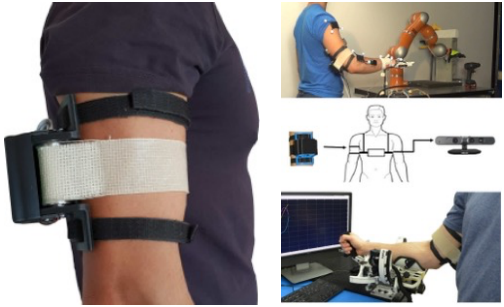

*Figure 1 The CUFF device applied on a subject's arm (on the left) and three possible application (on the right): from the top, in a teleoperation settings, as assistive device in a guidance system, and as assistive device for rehabilitation.*

In addition to tangential and directional cues, pressure and force information also represents another key informative cue for task accomplishment, e.g. in prosthetics to convey grip information to amputees [18]. In [19] a grip pressure feedback for a myoelectrical controlled prosthetic arm was proposed. A servo-controlled "pusher" was mounted to the socket and pressed into the skin an amount proportional to the force in the terminal device. Such a method was named "extended physiologic taction" (EPT). For this type of application, it is also important to minimize the need for additional sensors for grip force estimation, to likely increase users' acceptance.

To the best of authors' knowledge, there is little or no evidence of WHS able to deliver (i) normal force and (ii) tangential force at the same time. The existing devices in the literature act mainly at the finger level, and there are no existing systems that can fulfil requirements (i) and (ii) at the arm or forearm level. In [20], [21], authors presented a WHS that consists of two motors to convey both normal and tan- gential forces, thus simulating weight sensation. Analogously, a device - namely hring - with two servo motors that move a belt placed in contact with the user's finger skin can provide both normal and shear force to the skin in [22]. With this as motivation, we present a wearable device, the Clenching Upper-limb Force Feedback device, hereinafter referred to as CUFF (Figure 1), based on our preliminary work presented in the conference paper [23]. The CUFF consists of a fabric that can be worn by the users on their forearm.s The device is endowed with two independently controlled DC motors, thus enabling it to deliver both pressure and tangential cues in a fully wearable manner.

Compared with [23] in this work, we completely redesign the device to further increase its wearability and provide a thorough characterization of the system. Psychophysical experiments aiming to depict the perceptual workspace that the device can act on to elicit pressure and tangential skin stretch perceptions are also reported. Furthermore, a series of exploratory experiments are performed to assess the usability of the device in combination with an underactuated robotic hand.

The paper is organized as follows: Section II described the Motivation and the idea, while in section III the CUFF device will be described, showing all the mechanical features. In Section IV and V, a step by step description of the Characterization procedure (Physical and Psychophysical) is reported. Section VI present all the experiments performed with able-bodied participants. Sections VII and VIII present a discussion of the results obtained, conclusion and future works.

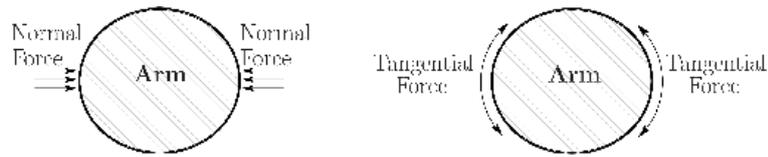

*Figure 2 Scheme of a arm, in section, with a distribution of normal pressure (left) and a distribution of tangential stress (right).*

# Section II: Motivation and Background

The problem of simultaneous rendering of multiple co- located stimuli over the same patch of skin is at the core of the design of cutaneous haptic devices. In wearable haptics this is especially important since there is a limited amount of devices that can be placed on a user, because of space and weight considerations.

Looking at the corpuscles that contribute to various tactile modalities of the skin, they respond to a very broad range of different stimuli, including temperature variation, vibration, pressure, stretch, etc. In our design ,we limit our analysis to the sensations of pressure and tangential stretch, as illustrated in Figure 2. This limitation comes from the consideration that:

- skin stretch can easily be associated with directional and navigation information [24];
- pressure cues can be used to successfully convey infor- mation on grip force for artificial limbs [25]; this is of obvious relevance in prosthetics, but also in tele-operation [26] and guidance scenarios [5];

Skin stretch has been investigated as a way to convey propri- oception feedback. There is evidence that the skin stretch is an important part of the mechanism that conveys information regarding proprioception of the hand and guidance informa- tion. In this sense, using skin stretch to convey proprioception of the prosthetic hands could lead to easier training for upper limb prosthesis users, since the feedback provided is felt in a way that is similar to a natural mechanism found in able- bodied limbs. Indeed, this type of feedback was shown to be an effective way to convey proprioception. In [14] rotational skin stretch was proposed as an alternative to vibrotactile feedback for conveying proprioception and results showed it to be more effective.

Regarding the second point, we were inspired by the pros- thesis and blind people world. A common requirement from the amputees are able to operate prostheses without constant visual attention and proprioception and force feedback has been shown to improve targeting accuracy under non-sighted conditions [27]. Furthermore, non-autonomous blind subjects are usually guided by the accompanying person with a hand placed on the arm, that conveys directional information to turn left or right with the skin stretch of the arm to the left or to the right.

In [23] we decided to focus on the skin surrounding the long trunks of the four limbs. This choice was motivated by the necessity of leaving hands and fingertips free of encumbrance and goes in the same direction often followed by customer electronics, where there are many examples of portable sys- tems which are designed to be strapped to the arm, the forearm, the thigh or the calf, such as music players, smartwatches, cellular phones, sphygmomanometers, fitness trackers etc. In this paper we will refine the design improving the previous work to obtain a device that is much more compact and lightweight, with a broader more in depth characterization of the system and its application possibilities.

# Section III: The CUFF device

The device is the result of an iterative design process in order to obtain a wearable device light and with small dimensions.

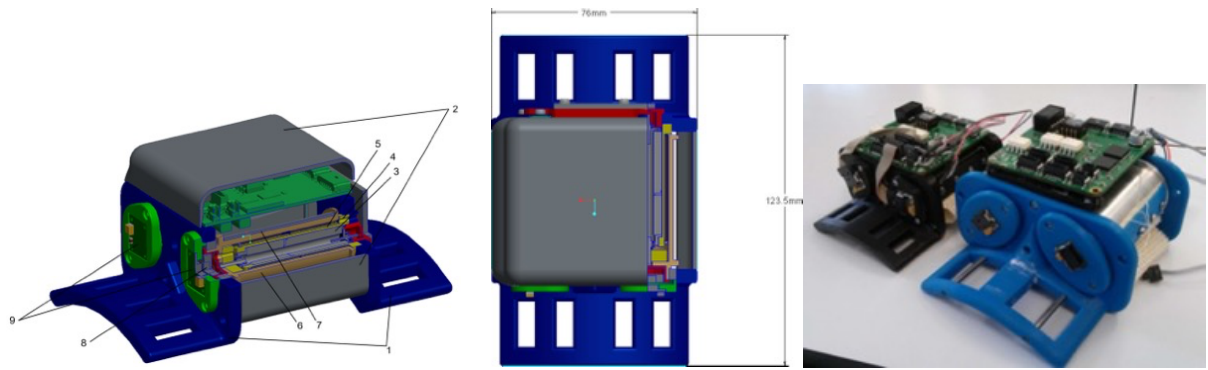

*Figure 3 Mechanical structure with all the components specified (a) and the size of the new prototype (b). In the figure is showed also the comparison between the old prototype, the one on the right side, and the new one on the left(c). In (a) it is possible observe the 1) main frame, 2) cover, 3) Motor, 4) Motor custom housing, 5) Needle bearings, 6) Cylinders, 7) Fabric locker, 8) Encoder magnet and 9)Encoder.*

## A. Design improvement

Starting from the characterization of the first prototype [23], we know that the belt can elicit a maximum normal force of 22 N with a total current absorption equal to 1000 mA. It is worth noticing that the current absorption is the sum of the current absorbed by the two motors. So the first step is to study the force perceived by the subjects with the first prototype. We conducted an experimental campaign with 6 subjects (2 female, mean age $26 \pm 2.1$) with different arm sizes ($28.90 \pm 25.6$ mm of circumference, measured with a tailor tape). The purpose of this experiment was to estimate the maximum force we could exert on the user's arm, before reaching an unpleasant level. We defined the unpleasantness level as the level where the subjects feel the arm squeezed but without feeling pain. During the experiment, the CUFF was worn on the participant's arm as in Figure 1. The motors were controlled to move in opposite directions with a step profile; every 2 seconds we increased the current absorption of 20 mA until the motor reached the stall condition. During the experiment we recorded the current absorbed by the motors and the force elicited. The outcomes of the characterization showed that the highest force value that we could exert was 17 N, which correspond to a current absorption of 800 mA.

Keeping in mind these results, the actual motors were chosen considering: the performed analysis, the diameter, and the reduction factor. The diameter was the main characteristic for dimension reduction, regarding the performed analysis we had a specific force value that had to be satisfied by the motor torque, and the reduction factor was chosen concerning the previous two characteristics and to the length of the gear itself.

In Figure 3 a render of the device (3(a) and 3(b)) is shown, highlighting some details of its mechanical implementation. This release has a mass equal to $\approx$ 230 g, while its overall dimensions are 124 x 70 x 58mm (compared with the original design had a mass of $\approx$ 494 g and overall dimensions 145 x 97 x 116 mm – see Figure 3(c)). With the new design the volume of the device reduced of about 55% and the weight of about 56% (see Figure 3(c)).

## B. Mechanical Description

Figure 3(a) shows the device, which can be considered composed of three main subsystems: a structural frame (com- prising parts 1, 2), the mechanical actuation units (3, 4, 5, 6) and the feedback interface (composed of a fabric belt not reported in the section view but visible in Figure 1(a)). Each actuation unit is powered by a DCX10L (by Maxon Motor) with a two stage planetary gear-head with a 64:1 gear reduction. The motors (3) are attached to the main frame (1) through a custom housing (4), which is fixed with a set of screws to the main frame; this is the fixed part of the actuation unit. A cylinder (6) is placed around the inner frame (4) and moves on it thanks to a set of needle bearings (5). This cylinder can rotate, actuated by the motor, around its main axis. During the movement, the cylinder is kept in axis through a bearing fixed to the main frame. Each actuation unit is sensed with a magnetic position sensor composed by the magnet (8), and its electronic board (9), (see Section III-C). The fabric belt, used as main feedback interface with the human body, is fixed to the cylinders through fabric fixing plates. The belt is a band of non-elastic fabric (0.4mm thick cotton 11-count aida cloth - 10 squares per inch), covered on its internal side with a silicone layer, to allow a better grip on the user's skin (biphasic bio- compatible silicone C-MOLAK8 by Colorificio Rodoero SAS, Arenzano (GE), Italy). The device is fastened on the user's arm through a Velcro band.

## C. Electronics and Software

The CUFF is controlled using a custom made electronic board (PSoC-based electronic board with RS485 communica- tion protocol), directly applied on the device. The board can be powered from 5V up to 24V; all external logic ports are 3.3V. The used magnetic encoders are AS5045 Rotary Sensor (by Austria Microsystem), with angular resolution is about 0.0879 corresponding to 4096 ticks per revolution.

The CUFF is capable of working as a standalone device directly connected to the SHP (as force, proprioception or combined force-proprioception feedback), but it can also be connected to a computer and can be controlled using its software or its C++ and Matlab© libraries freely available online. This allows a simple and fast integration in a wide group of applications. All softwares and more information can be freely downloaded from Natural Machine Motion Initiative website[1][28].

---

[1] www.naturalmachinemotioninitiative.com

# Section IV: Device Characterization

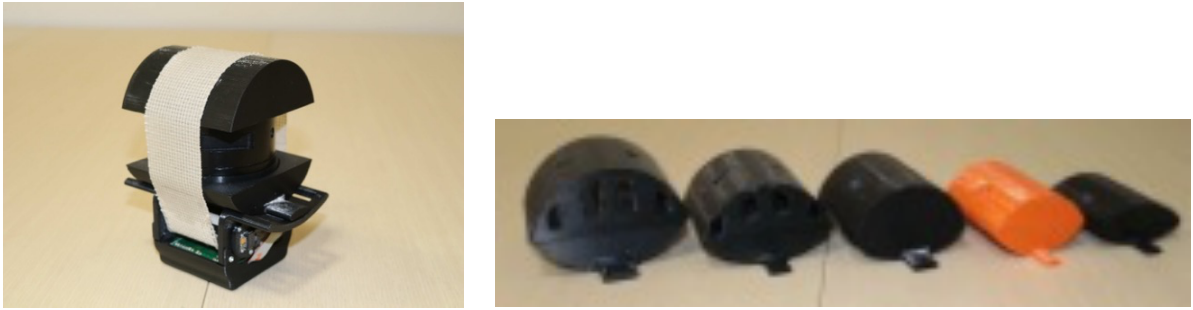

*Figure 4 Force sensor structure for characterization procedure mounted on the CUFF (a) and all the 3D-printed structures with the different radii (b).*

To have a better knowledge of the device, we performed: a physical characterization (see section IV-A)of the new device, and a psychophysical characterization (see section IV-B) of the tangential displacement and normal force just noticeable difference (JND).

## A. Physical Characterization

The goals of the process were: 1) to obtain a relation between the position of the motor and the normal force exerted by the fabric belt on the subject's arm, (concerning a variable zero position), and 2) to determine the relationship between the exerted force and current absorbed by the motors. A 3- axis force sensor, the ATI Gamma, was used to measure the normal force exerted on the subject's arm. We 3D printed structures with different radii to reproduce various human arm sizes. Starting from different zero positions, five cylinders were designed with the following dimension (80, 85, 90, 100, and 115 mm). Figure 4(a), and figure 4(b) show the final structure used for the characterization and the cylinders created, respectively.

### 1) Procedure and Control Mode

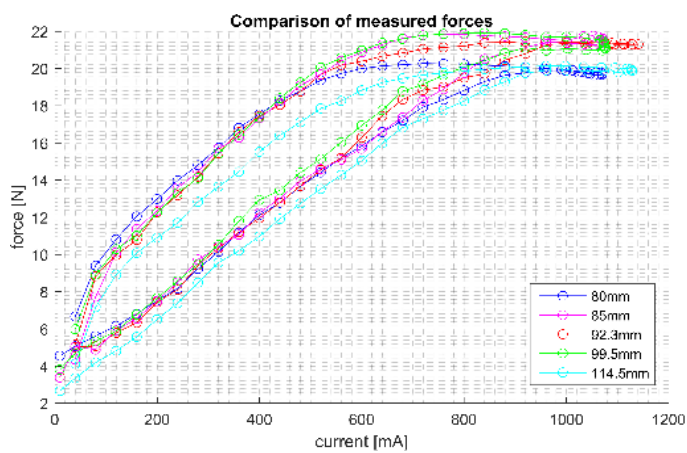

*Figure 5 Characterization curve obtained with non-elastic fabric band and current control.*

The characterization was performed using a series of steps of tighten-release movements of the CUFF. More specifically, the two motors were controlled to move in opposite direction to pull the fabric without generating side-scrolling movements. Ten repetitions per radius were performed. At the beginning of any repetition, an auto-adjustment (hereinafter referred with pre-tensioning phase) of the fabric around the cylinder was performed. During this phase the fabric was tightened up to the stall condition of the two motors and then released up until the absorbed current was close to zero, guaranteeing the complete adhesion on the lateral surface of the cylinder. the CUFF on the sensor was 0.464 N ±0.1793 N. This provides a zero reference starting position for all the characterization procedures. After the pre-tensioning phase, the tighten-release movement was performed. We tested two control strategies: a current control, and a position-current control. In the current control we commanded a maximum current absorption of 1000 mA, while for the position-current control a maximum position of 800 motor ticks (corresponding to 17). It is worth noticing that the position-current control consist of a double control loop (see fig 6). The position controller generates the reference current, and the current controller generates the PWM output, which feeds the motor to generate the movement. This controller structure enables a double error check, one for the current and another one for the position. During the tighten phase, the maximum value was reached in 60 seconds, and then the direction was reversed for other 60 seconds back to the starting position. We recorded the current, the position of the motors and the normal forces exerted on the sensor.

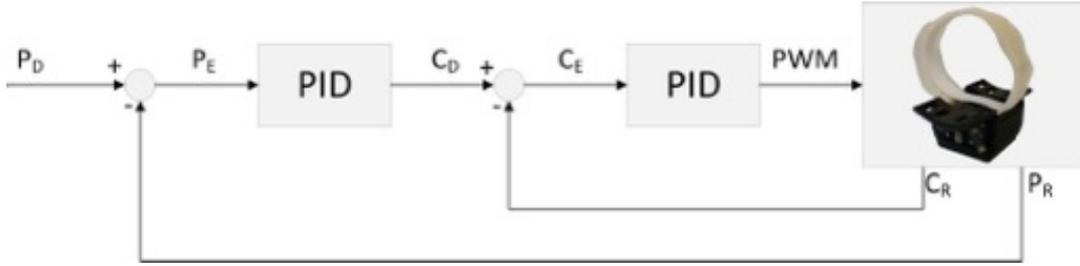

*Figure 6 CUFF double control loop - External position loop and internal current loop.*

The control current showed a too wide hysteris effect between the tighten-released phase (see Fig. 5), with respect to the position-current control (Fig. 7(a)). For this reason, we decided to used the CUFF in position-current modality for psy- chophysical characterization and the experimental campaign with users.

*2) Data Validation*

To confirm the goodness of the data acquired with the position-current control. We are interested in a function that given a desired force as input returns the cor- responding position command to the CUFF. We found that the function that provides the best fit for the data set is a 3rd order polynomial function

$$P_D = f(F_D) = 0.1138F_D^3 - 5.204F_D^2 + 89.22F_D$$

Where $P_D$ is the desired motor position and $F_D$ is the desired force (Adjusted R-square of 0.9335). We tested the correctness of the fitting (see Fig. 7(b)) in another series of experiments, using the function $f(F_D)$ to get the position values to command the CUFF and measure the forces we obtained. We get an RMSE between theoretical force values and measured ones of 1.32 N considering all the experimental trials for all the radii (see Figure 8).

## B. Psychophysical Characterization

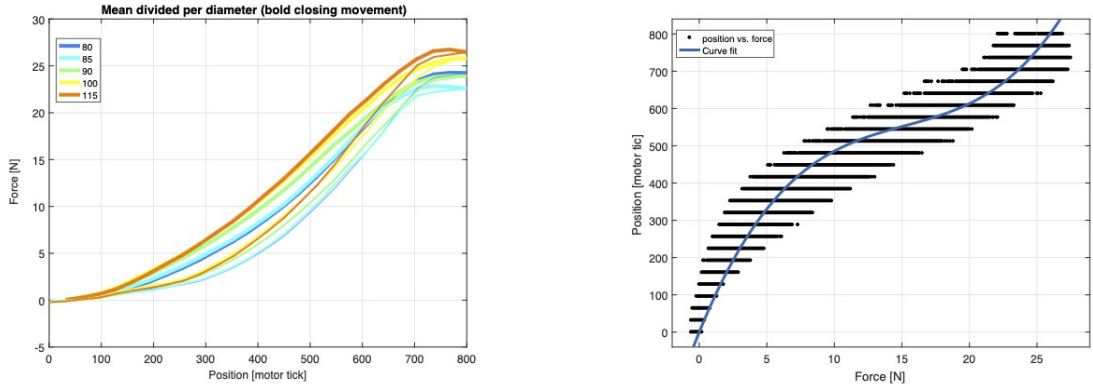

*Figure 7 In (a) the characterization data, obtained with the position-current control modality are showed. The plot reported the mean values for each diameter for the tighten movement (thicker lines) and releasing one. (b) shows the data set (in black) and the obtain fitting curve (in blue).*

The goal of the Psychophycal characterization was to es- timate user's ability to discriminate the normal force and the tangential displacement delivered with the CUFF. We used the method of the constant stimuli to find the Just Noticeable Difference (JND), as defined in [29], i.e., the minimum amount of displacement/force that can elicit a different perception in users with respect to a reference stimulus. We considered to use the device in the two operating modes depicted in Figure 2. Eleven right-handed able-bodied participants (7 females, mean age ± SD: 26.64±8.86) gave their informed consent to participate in the experiment. No one had any physical limita- tion which would have affected the experimental outcomes. The average arm diameter (using a cylindrical geometrical approximation) for participants was 80.3 ± 7.7 mm (mean ± SD).

### 1) Tangential Displacement Perception

*1) Stimulus and Procedure*: Participants were comfortably seated, with their forearm resting on a desk and, wearing the cuff on their arm, see Figure 1. During the experiments, subjects wore headphones with white noise to prevent auditory cues; in addition, they were required to not look at the device during the experiments. Before these experiments, the CUFF was applied to the arm of the subject and we performed a pre-tensioning and rescaling procedure. The pre-tension was performed to bring the fabric in contact with the subject's skin as previously described with the sensorized cylinders. The new position of the motors was used as zero reference position for all the trials. The rescaling was performed to rescale the positions obtained with the fitting function to the arm biomechanical structure. More specifically, the rescaling was computed commanding to the motor a position much higher than the maximum one that can be effectively reached by the CUFF motors (due to the current limits). We know that the maximum reachable position corresponding to the maximum current absorption provides on average a force of 25 N on the sensor. The effective position of CUFF motors obtained in this way was then associated with the value of 25 N and all the other points of the characterization were re- scaled accordingly. Participants received paired stimuli, each stimulus was related to the tangential displacement provide by the rotation of the CUFF's motors in the same direction, and were asked to indicate which stimulus in pairs was perceive as higher. Each pair consisted of a reference stimulus (RS) and a comparison stimulus (CS), presented in a pseudo- random order. The displacement was equal to 17.91 mm in the reference stimulus and was pseudo-randomly chosen among five discrete and equally spaced values (corresponding to 5.97, 11.94, 17.91, 23.88, and 29.85 mm) in the comparison stimulus. A single trial consisted of : the first stimulus, an inter-stimulus interval of 2 s, and the second stimulus followed by the subject's response. We considered two experimental blocks corresponding to the direction of the tangential motion (i.e. rightward or leftward). Each block consisted of 100 trials. The binary responses of the participants were recorded.

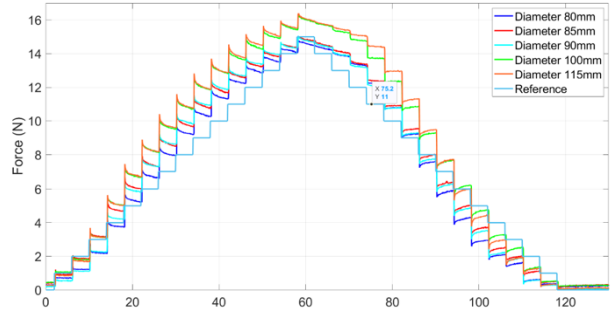

*Figure 8 Validation experiment — In orange the theoretical force values, while the other colors refer to the measured force values for different cylinder radii.*

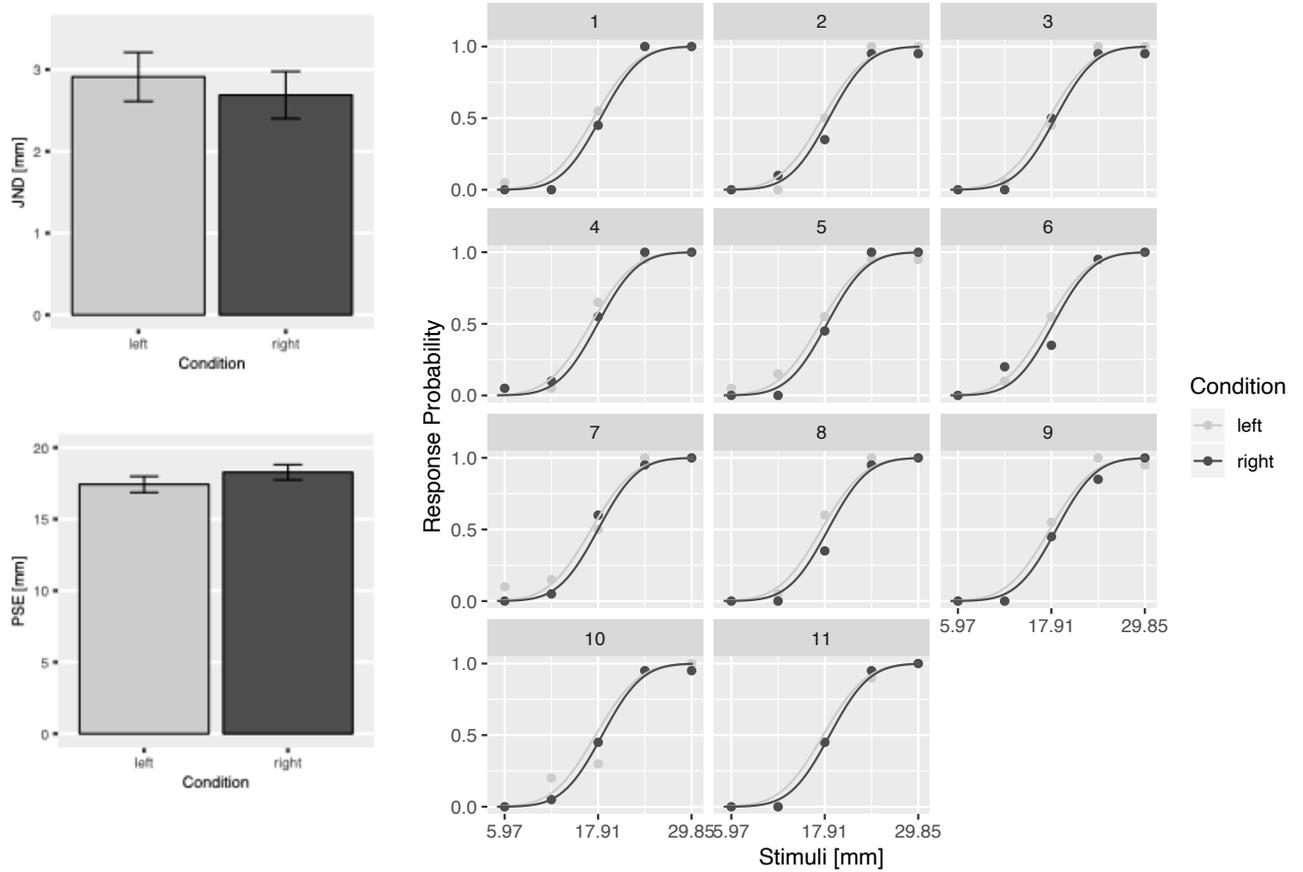

*Figure 9 On the left the JND and PSE bar plots for the two conditions. On the right the raw data and GLMM predictions for right and left condition are presented, for all the subjects. Each box represents raw data and model predictions for each single participant (labeled as 1-11)*

*2) Data Analysis*: We modelled the responses of each volunteer using the psychometric function

(2) $$\Phi^{-1}\big[P(Y_j = 1)\big] \sim \beta_0 + \beta_1 x_j$$

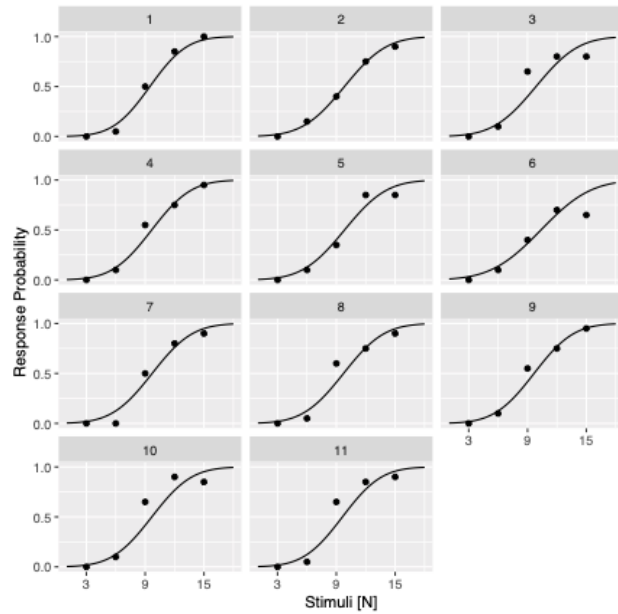

*Figure 10 GLMM fit for all the subjects. Each box represents raw data and model predictions for each single participant (labelled as 1 - 11).*

where [P (Yj = 1)] is the probability that, in trial j, the participant reported a larger stimulus in the comparison than in the reference stimulus, 1 is the probit transformation of the response probability (i.e., the inverse function of the cumulative normal distribution), and $x_j$ is the value of the comparison stimulus. We extended the psychometric function to the whole pool of participants through a Generalized Linear Mixed Model (GLMM) [30], which enables to take into account the analysis of clustered data. In our case a cluster is the collection of re- peated responses in several participants. GLMM is a hierarchi- cal model that provides both predictions on the experimental effects, which are assumed to be systematic across participants, and an estimate of the variability between participants. For each experimental condition, we estimated the Just Noticeable Difference (JND), i.e. the amount of stimulus change for a difference to be noticeable and the Point of Subjective Equality (PSE), i.e. the stimulus value yielding a response probability of 0.5. The JND provides an estimate of the precision of the response: the lower is the JND, the easier is to discriminate two stimuli of different lengths. The PSE estimate the accuracy of the response: the response is accurate if the PSE is non-significantly different from the reference stimulus. For each parameter, we estimated the 95% confidence intervals (CIs) using the bootstrap method described in [30].

*3) Results*: The GLMM fit to the data is illustrated in Figure 9(c). The JND was equal to 2.91 mm (95% CIs: 2.61–3.21 mm) for rightward moving stimuli and 2.70 mm (95% CIs: 2.40–2.96 mm) for leftward moving stimuli. The difference in JND between the two conditions was not statistically significant (difference: 0.22 mm; 95% CIs of the difference from -0.19 to 0.63 mm). The PSE was equal to 17.42 mm (95% CIs: 16.89–17.95 mm) for rightward moving stimuli and 18.28 mm (95% CIs: 17.76–18.83 mm) for leftward moving stimuli. This means that the response was accurate for rightward moving stimuli, and slightly biased for leftwards (although the bias was smaller than 1 mm). The 95% CI of the difference between PSEs does not include zero (difference: -0.85 mm; 95% CIs of the difference from -1.54 to 0.18 mm). This small bias must be further investigated. A possible explanation could be related to the non-linearity of human arm biomechanics.

*2) Normal Force Perception*

*1) Stimulus and Procedure* The experimental setup and procedure were the same as for the tangential displacement task. Participants experienced paired stimuli and they were asked to indicate which stimulus in the pair produced a higher normal force. The force values were equal to 9 N in the RS and was pseudo-randomly chosen among five stimuli (3, 6, 9, 12, 15 N) in CS. The stimulus duration was 1 s, while the inter-stimulus interval was 2 s.

*3) Data Analysis*

Data were analyzed with modalities similar to the one of tangential displacement task.

*4) Results*

The estimated JND was 2.21 N (95% CIs: 1.95–2.47 N). The estimated PSE was 9.75 N (95% CIs: 9.38–10.11 mm). That is, the responses were precise (the participants were able to discriminate a force difference as small as 2 N). As for the accuracy, there was a small difference between the PSE and the reference stimulus (smaller than 1 N). In Figure 10 it is possible to see the GLMM fit.

# Section V: Explorative experiments with humans

A series of experiments were performed to test the device in two of its possible applications, as wearable haptic feedback effects, which are assumed to be systematic across participants, and an estimate of the variability between participants. For each experimental condition, we estimated the Just Noticeable Difference (JND), i.e. the amount of stimulus change for a difference to be noticeable and the Point of Subjective Equality (PSE), i.e. the stimulus value yielding a response probability of 0.5. The JND provides an estimate of the precision of the response: the lower is the JND, the easier is to discriminate two stimuli of different lengths. The PSE estimate the accuracy of the response: the response is accurate if the PSE is non-significantly different from the reference stimulus. For each parameter, we estimated the 95% confidence intervals (CIs) using the bootstrap method described in [30].

*3) Results*: The GLMM fit to the data is illustrated in Figure 9(c). The JND was equal to 2.91 mm (95% CIs: 2.61–3.21 mm) for rightward moving stimuli and 2.70 mm (95% CIs: 2.40–2.96 mm) for leftward moving stimuli. The difference in JND between the two conditions was not statistically signif-device for prosthetic applications, and for remote control of avatars. The two possible degrees of freedom of the device can be fully used in this application, enabling the generation of different stimuli connected to different perceptions. Other examples of applications of the CUFF are discussed in Section VI.

For this series of experiments, the CUFF was used with the Pisa/IIT SoftHand. The CUFF is designed to be able to work with the SoftHand as a standalone device, managing all the communications between the two systems. No PC is needed for the communication.

Three different feedback were tested: proprioception, force and proprioception plus force feedback. The combination of the two feedback was tested to assess if the device was able to produce multimodal feedback.

*A. Modality Control*

As previously said, we decide to use the CUFF device in conjunction with the SoftHand. We identified three stimuli, that the feedback could generate: a skin stretch for the proprioception feedback, a squeeze for the force feedback, and a union of skin stretch and squeeze for the combined information of proprioception and force. Regarding the pro- proioception feedback, the measured position of the motor by the encoder on the SoftHand, pSHmeas, is mapped in the reference position for the motor's CUFF, pCref . Every closure of the hand produce a rotation of the motors in the same direction (conventionally chose to the left), generating a skin stretch stimulus. Instead, concerning the force feedback, the residual current of the SoftHand, rcSHmeas, is mapped into opposite reference position for the CUFF motors, rcCref, generating a squeeze stimulus on the user's arm. It is worthy notice that the residual current is the difference between the estimated current and the real current absorbed by the SoftHand motor. Regarding the combined stimuli, we used the motor position and residual information coming from the SoftHand, to generate the combined movement of stretch and then squeeze of the arm. The tactile stimulation of the CUFF device directly depends on the information received from the SoftHand; in particular the encoder position give information that can be used to understand whether the hand is grasping an object, and so can give insight on the dimension of the object that is being grasped. From the residual current we can receive information regarding the softness of an object: more the residual is higher more the object, that the hand is squeezing, is hard; so it can give information about the fragility of the object.

We chose six cylinders made with different materials: three rigid cylinders, printed with the 3D printer in Rapid Prototyping material (ABS), and three soft cylinders made with foam rubber to use for the discrimination test of size, softness and the combined information (size plus softness). We selected three diameters (40, 60, 80 mm), and we recorded the pSHmeas and rcSHmeas value during the grasps phase, including the hand completely closed around the object. More specifically each cylinder was positioned in the palm of the SoftHand, horizontally, and it was grasped ten times. Indeed for the softness discrimination task, we used a cylinder of 60 mm of diameter with inside a 3-axis force sensor, the ATI Gamma, to measure the force elicited by the SoftHand on an object with three different closures. We chose a final closure position of the hand of 15000, 16500, and 1800 ticks. Also in this case, we recorded the residual current value, and the cylinder was positioned horizontally in the palm of the robotic hand; we repeated the procedure ten times for each closure.

The CUFF device receives the information directly from the SoftHand via RS485 protocol at a frequency range of 1 kHz. Mapping the information coming from the robotic hand into feedback device, means associate to each value of the hand a corresponding stimulus delivered by the device. At the begin- ning of our investigation, we chose to use a linear function to associate each value of the position and the residual current of the SoftHand with the reference position for the CUFF device. We used a linear mapping had two constrain: $p_{SHmeas} = 0$ corresponded to $p_{Cref} = 0$ and $p_{SHmeas} = p_{SHmax}$ to $p_{Cref} = p_{Cmax}$; where $p_{SHmax}$ is the maximum value of the SoftHand encoder and pCmax is the maximum value of the CUFF encoders. Same constrains for the linear mapping of the residual current: $rc_{SHmeas} = 0$ corresponded to $rc_{Cref} = 0$ and $rc_{SHmeas} = rc_{SHmax}$ to $rc_{Cref} = rc_{Cmax}$. Therefore, the reference position is computed as:

(3) $$p_{Cref} = \left(\frac{p_{SHmeas}}{p_{SHmax}}\right) p_{Cmax}$$

$$(4) \qquad rc_{Cref} = p_{SHmeas} * gain$$

respectively for the proprioception feedback (3), and for the force feedback (4), with a $gain = 0.4$ (heuristically found). After the first set of experiment (see sec V-B), we decided that a non-linear mapping could be a better solution for the discrimination task of size and force. So, for subsequent experiment, we decided to test an exponential mapping for the proprioception feedback, and logarithmic mapping for the force feedback. We used the same constrains used for the linear mapping, $p_{SHmeas} = 0$ corresponded to $p_{Cref} = 0$pCref = 0, but we imposed that $p_{SHmeas} = \frac{2}{3} p_{SHmax}$ was mapped to $p_{Cref} = \frac{1}{3} C_{max}$. So the new reference position was:

$$(5) \qquad p_{Cref} = \alpha e^{-\beta \left( \frac{p_{SHmeas}}{p_{SHmax}} \right)} p_{Cmax}$$

$$(6) \qquad rc_{Cref} = \gamma log \left( 1 - \delta \frac{rc_{SHmeas}}{rc_{SHmax}} \right) rc_{Cmax}$$

The constrains were imposed by choosing the correct values for the constants: $\alpha = 0.1547, \beta = 1.944, \gamma = 0.9510$ and $\delta = -0.3317$.

For the discrimination task of the combined information (size plus force) we used at the beginning the linear mapping, and then the exponential and logarithmic mapping for size and force respectively.

## B. Experimental Protocol

We tested the effectiveness of our system with a total of 32 participants. The main goal of these experiments was to verify if our feedback device could be a potential solution to translate informations about the size and the softness of an object in a proprioception and force stimuli. We investigated also, the possibility to give a combine information of propri- oception and force at the same time. None of the participants had any physical or mental limitation which could affect the experimental outcomes. We designed two experiments: a discrimination task and a fragile object task.

### 1) Discrimination task: Participants

Thirty-two able-bodied subjects all right-handed gave their informed consent to participate to the discrimination task. Sixteen participants (5 females, mean age ± SD: 27.68 ± 2.6), perform the task of testing a linear mapping between the CUFF and the SoftHand for all the modalities. Sixteen subjects (4 Female, mean age ± SD: 28.87 ± 2.6), accomplished the task testing a logarithmic mapping for the grasping force modality, exponential mapping for the proprioception modality. Both the mappings were used for the combined information modality.

### 2) Discrimination task: Setup and Procedure

The experiment was composed of a training phase and three discrimination tasks: (1) proprioception modality, (2) grasping force modality, and (3) combined information modal- ity. During the experiment, the participants were comfortably seated wearing the CUFF on the right arm, and with the forearm placed on the table. They were blindfolded and wore headphones with pink noise to prevent the usage of any auditory cue generated by the rotation of CUFF motors. In all the discrimination task, the subjects perceived on their arm the stimuli produced by the feedback device in a playback modality. We used the recorded signals from the SoftHand, during the grasping phase of the cylinders (see sec V-A), to control the rotation of the motor's CUFF. It is worth noticing, that we decided to adoptthis strategy in order to have the subjects completely focused on the stimuli elicited by the device. During the training period of 15 minutes subjects saw the dimension and material of the cylinders, and they tried to associate the stimulus generated by the CUFF with the corresponding cylinders. The order of execution for the proprioception and grasping force modality was randomized and counterbalanced between the subjects. The combined information modality was always conducted as third task to al- low the subjects to first have a clear idea of the proprioception and force stimulus individually.

*Proprioception Feedback*

We asked participants to recognize the sliding stimuli as- sociate to the size of three rigid cylinders plus the stimulus associate to the closure of the hand without object (i.e close hand without object, ∅40 mm, ∅60 mm, ∅80 mm). During each trial, the subjects perceived on their arm, a sliding movement of CUFF's belt to the left. The belt would move less, when the cylinder grasped by the SoftHand is of larger size, on the other hand the belt would moved more when the SoftHand closed around small or without any object. Subjects were presented sixteen pairs of stimuli, and asked to tell whether the second stimulus was smaller, bigger or of equal sliding with respect to the first stimulus. To avoid artefacts in the results due to time error, each pair of stimuli was presented two times, one with the first element of the pair being presented as first and another one with the first element of the pair presented as second in temporal order, for a total of twenty trials in random order.

*Force Feedback*

During this task, the participants had to recognize the intensity level of three squeeze stimuli plus the stimulus originated by the closing of the SoftHand without the object. During each trial, the subject perceived a squeeze stimulus elicit by the belt around their arm. The belt would squeeze the arm more, when the SoftHand grasped a rigid object and less when the SoftHand grasp a soft object. For this modality, we decided to use the residual current values obtain during the recording session of the sensorized cylinder (close hand without object, 15000 ticks, 16500 ticks, 18000 ticks). Also in this case, subjects were presented sixteen pairs of grasp stimuli and asked to tell if the second stimulus was bigger, smaller or of equal intensity with respect to the first stimulus. Each pair of stimuli were presented two times, as in the proprioception modality, for a total of twenty trials in random order.

*Combined Feedback*

We asked participants to recognize the combined stimuli of proprioception plus grasping force. The participants perceived first the sliding of the belt to the left, and immediately after the squeeze of the arm. With these two information the participants should be able to distinguish size and softness of the cylinder. As said in the previous modality, the belt would move less, if the hand grasp a big size cylinder, and would squeeze harder when the hand grasp a rigid cylinder. For this task, we used the signals recorded from the two sets of cylinder (i.e. close hand without object, ⊣40 mm, ⊣60 mm, ⊣80 mm rigid and soft.) Subjects were presented the forty- nine pair of combined stimuli and ask to distinguish if the second stimulus was bigger, smaller or of equal sliding and intensity with respect to the first one. Also in this modality each pair was presented two times, to avoid artefacts in the results, for a total of fifty-six trials presented in random order.

### 3) Fragile Object Task: Participants

Sixteen subjects (4 Female, mean age ± SD: 28.87 ± 2.6), all right-handed, gave their informed consent to participate to the Fragile Objacet Task.

### 4) Fragile Object Task: Setup and Procedure

We decided to add another task, where the subjects con- trolled the SoftHand via EMG sensors. They wore the CUFF device on their arm and the EMG sensors were positioned on the Flexor Digitorum Superficialis (FDS) and on the Extensor Digitorum Communis (EDC), and held in place by an elastic band. The SoftHand was connected to a gravity compensator, the SaeboMAS, to help the subjects to support the arm during the duration of the experiment. The participants were standing in front of a desk, with the arm placed on the gravity compen- sator, wore headphones with pink noise to avoid auditory cue, and glasses with covered lenses to not allow a clear view of the boxes. We asked participants to accomplish a task similar to a Box and Block test: in one minute they had to move as much as possible fragile objects from a point A to a point B overcoming an obstacle, without breaking them. This task was used in [31] to assess the grsping efficiency. The subjects performed the task in four modalities: (1) no feedback, (2) proprioception feedback, (3) force feedback and, (4) combined information feedback. Each participant repeated the task four times, one for each modality. The order of modality execution was randomized.

The fragile object is a box made of paper, dimension 50x50 mm, with two lateral sides open and, a fragile fuse placed in the middle of the other two side. The fragile fuse is a piece of pasta, enough fragile to be broken with a pinch grasp. We tested three types of spaghetti with a diameter of 0.11, 0.15 and, 0.2 mm, and we characterized a piece of spaghetto 50 mm long, to compression with an Instron instrument. We repeated the test 15 times for each diameter, and recorded the force corresponding to the breaking point.

For the experiment, we chose the spaghetto with a diameter of 0.15 mm with a breaking force approx equal to 35 N.

## C. Data Collection

To evaluate the performance in the discrimination task we recorded: the number of correct answers for all the discrim- ination tasks (see Fig 11), and the results of a subjective quantitative evaluation performed by administrating a Likert scale survey. The Likert scale consists of questions about the system and the experimental task, to which participants had to answer by assigning a score ranging from 1 *totally disagree* to 7 *totally agree*. This represents a common procedure to evaluate devices for assistive robotics and Human-Robot Interaction [32]. The questions delivered to the subjects are listed in Table I. Questions from Q1 to Q21 were responded to all the subjects of the discrimination task. Questions from Q22 to Q38

| | Questions | Participants Experiment A | | Participants Experiment B | |
|---|---|---|---|---|---|
| | | Mean | Std. Dev. | Mean | Std. Dev. |
| Q1 | I felt hampered by the cutaneous device | 2.18 | 1.22 | 1.75 | 0.85 |
| Q2 | The belt produced pain during long time usage | 2.18 | 1.22 | 1.43 | 0.51 |
| Q3 | The noise generated by the actuators interfered with the haptic perception | 1.93 | 0.99 | 1.31 | 0.6 |
| Q4 | The sensation provided by the CUFF on the arm felt pleasant | 4.68 | 1.40 | 4.81 | 0.98 |
| Q5 | The sensation provided by the CUFF on the arm felt unpleasant | 2.25 | 1.48 | 2.68 | 1.07 |
| Q6 | During the discrimination task the CUFF and the cylinder were out of my visual field | 6.56 | 1.5 | 6.81 | 0.4 |
| Q7 | During the discrimination task, I was able to see the cylinder or the CUFF device | 1.5 | 1.5 | 1.12 | 0.34 |
| Q8 | During the discrimination task, I was able to hear the sounds made by the motors of the CUFF | 2.56 | 2.3 | 1 | 0 |
| Q9 | It has been easy to discriminate the cylinder size using the CUFF in proprioception modality | 4.93 | 1.12 | 5.75 | 0.57 |
| Q10 | Discriminating the cylinder size without looking at them was very difficult using the CUFF | 2.81 | 1.47 | 2.31 | 0.6 |
| Q11 | The proprioception feedback (sliding movement) was clear | 5.62 | 1.08 | 5.93 | 0.68 |
| Q12 | I was able to understand the position of the hand also after a distraction, perception retention | 5.62 | 1.08 | 4.25 | 1.61 |
| Q13 | It has been easy to discriminate the cylinder softness using the CUFF in force modality | 4.93 | 0.77 | 5.43 | 0.62 |
| Q14 | Discriminating the cylinder softness without looking at them was very difficult using the CUFF | 3.12 | 1.58 | 2.31 | 0.7 |
| Q15 | I was able to distinguish different levels of force through the cutaneous device. | 5.43 | 1.03 | 5.75 | 0.68 |
| Q16 | The maximum force elicit pain. | 1.43 | 0.72 | 1.18 | 0.54 |
| Q17 | It has been easy to discriminate the cylinder softness using the CUFF in combined modality. | 5.75 | 0.77 | 5.37 | 1.14 |
| Q18 | It has been easy to discriminate the cylinder size using the CUFF in combined modality. | 5.62 | 0.71 | 4.8 | 0.91 |
| Q19 | Discriminate the cylinder size without looking at them was very difficult using the CUFF. | 2.81 | 1.42 | 2.31 | 0.79 |
| Q20 | Discriminate the cylinder softness without looking at them was very difficult using the CUFF. | 2.68 | 1.44 | 2.62 | 0.61 |
| Q21 | The two stimuli were clearly distinguishable. | 5.93 | 0.77 | 5.06 | 0.85 |
| Q22 | It has been easy to use the SoftHand together with the CUFF. | / | / | 6.06 | 0.68 |
| Q23 | I was feeling uncomfortable while using the SoftHand together with the CUFF. | / | / | 1.87 | 0.8 |
| Q24 | It was easy to move the virtual eggs while receiving the proprioception feedback by the CUFF. | / | / | 5.75 | 1.06 |
| Q25 | It was easy to move the virtual eggs while receiving the force feedback by the CUFF. | / | / | 5.87 | 0.61 |
| Q26 | It was easy to move the virtual eggs while receiving the proprioception feedback by the CUFF during the combined modality. | / | / | 5.68 | 1.25 |
| Q27 | It was easy to move the virtual eggs while receiving the force feedback by the CUFF during the combined modality. | / | / | 5.93 | 0.68 |
| Q28 | I had feeling of performing better while receiving force feedback or proprioception feedback by the CUFF with respect to no feedback. | / | / | 6.18 | 0.75 |
| Q29 | I had the feeling of performing better when I was not receiving any feedback by the CUFF. | / | / | 1.87 | 0.71 |
| Q30 | It has been easy to discriminate the cylinder size without any feedback. | / | / | 1.06 | 0.25 |
| Q31 | It has been very difficult to discriminate the cylinder size without any feedback. | / | / | 6.87 | 0.34 |
| Q32 | It has been easy to discriminate the cylinder softness without any feedback. | / | / | 1.37 | 0.5 |
| Q33 | It has been very difficult to discriminate the cylinder softness without any feedback. | / | / | 6.93 | 0.25 |
| Q34 | It has been easy to discriminate the cylinder size without any feedback during the combined modality. | / | / | 1.18 | 0.4 |
| Q35 | It has been very difficult to discriminate the cylinder size without any feedback during the combined modality. | / | / | 6.93 | 0.25 |
| Q36 | It has been easy to discriminate the cylinder softness without any feedback during the combined modality. | / | / | 1.18 | 0.4 |
| Q37 | It has been very difficult to discriminate the cylinder softness without any feedback during the combined modality. | / | / | 6.75 | 0.44 |
| Q38 | At the end of the experiment I felt tired. | / | / | 2.56 | 1.41 |

TABLE 1: These statements were rated by the subjects of Experiment A and B using a 7-point Likert scale (1: Strongly disagree, 7: Strongly agree). Means and standard deviations across all individuals are reported.

| | No Feedback | | Proprioception | | Force | | Combined stimuli | |
|---|---|---|---|---|---|---|---|---|
| | Mean | Std. Dev. | Mean | Std. Dev. | Mean | Std. Dev. | Mean | Std. Dev. |
| Broken egg | 0.87 | 1.2 | 0.18 | 0.4 | 0.62 | 0.95 | 0.56 | 0.89 |
| Regrip | 1 | 1.09 | 0.56 | 0.72 | 0.68 | 0.87 | 0.56 | 0.89 |
| Wrong grasp | 1.37 | 1.14 | 0.56 | 0.69 | 0.68 | 1.3 | 0.43 | 0.72 |
| Good grasp | 9 | 3.65 | 9.5 | 2.8 | 9.06 | 3.29 | 9.37 | 4.44 |
| Total egg | 10.37 | 3.82 | 10.06 | 3.01 | 9.75 | 3.17 | 9.81 | 3.91 |

TABLE II: The table shows the mean and standard deviation to accomplish the virtual eggs tasks for each different experimental modality.

were delivered to the participants, who accomplished the virtual eggs task. In particular,

question from Q1 to Q5 referred to the wearability of the CUFF device, questions from Q6 to Q8 were related to visual and auditory distraction. Questions from Q9 to Q21 investigated the intuitiveness of sensory substitution, the sensation provided by the device and the performance. Questions from Q22 to Q38 were about the use of the CUFF device in conjunction with the SoftHand, and they investigate the performance, the usefulness of the stimuli and the fatigue during the task. Regarding the fragile object task, in addition to the Likert scale survey, we recorded for each modality of execution: the total number of eggs moved, the number of broken eggs, the number of regrip, and the wrong/good grasp (see Table II).

### D. Results

Starting with the discrimination task using the linear map- ping, we can see from fig 11(a), 11(c), that the proprioception and force feedback had a percentage of success above of 50%. Only in a few cases the results showed percentage below 50%. The combined information showed good results only for some specific pairs of the cylinder with percentage of success close by 100%. Instead of using the logarithmic and exponential mapping, the proprioception and the force feedback obtained a percentage of success above 50% for all the pairs presented to the subjects. Table I reports the Likert scale results for the experiments. The question Q1 to Q3 (related to the wearability of the device) showed a positive rate with a mean value of 2.18, 2.18, 1.93 and a standard deviation of 1.22, 1.22 and 0.99. The questions from Q6 to Q12, regarding the discrimination task with the proprioception feedback obtained a positive rate, especially question Q11 had a mean value of 5.62 and a standard deviation of 1.08 (for the linear mapping), and a mean value of 5.93, and a standard deviation 0.68 (for the logarithmic/exponential mapping) Questions from Q13 to Q16, investigated the force feedback discrimination task. The performance of the task, Q13 had a mean value of 4.93 and a standard deviation of 0.77 (linear mapping), and a mean and standard deviation value of 5.43 and 0.62 respectively (logarithmic/exponential mapping). It is worth noticing question Q16, regarding the maximum force elicited by the device, that presented a mean value of 1.43, 1.18 and a standard deviation of 0.72, 0.54 for linear and logaritmic mapping respectively. Questions from Q17 to Q21, regarding the combined stimuli feedback task, showed a positive rate. Subjects, who took part in the fragile object task, underwent also the questions from Q22 to Q38. For question Q22 and Q23, about the wearability and the use of the device in conjunction with the SoftHand, the mean values were 6.06, 1.87 and a standard deviation of 0.68 and 0.8 respectively. Questions from Q28 to Q29, about the performance, were rated with mean values of 6.18and1.87 and a standard deviation of 0.75 and 0.71. For questions from Q30 to Q38, about the usefulness of the stimuli elicit by the device, we received a positive rate. For question Q38, regarding the fatigue, mean and standard deviation values were 2.56 and 1.41. Table II shows the results of the virtual eggs task. On average the total eggs moved had a mean value of 9, 9.5, 9.06, 9.37, and a standard deviation of 3.65, 2.8, 3.29 and 4.44 for the no feedback, proprioception, force and combined stimuli modality respectively. The total number of broken eggs for the four conditions, on average, is less than 1.

# Section VI:  Discussion

The results of the psychophysical characterization and the experimental validation with able- bodied subjects demon- strated the effectiveness of the CUFF device to convey proprio- ception and force feeback stimuli. The tangential displacement demonstrated a non statistically significant difference between the leftward and rightward sliding stimulus. Furthermore, they were able to perceive a force difference as small as 2 N. The results of the discrimination task using the linear mapping, showed that the participants were able to correctly recognize the pairs of the cylinders, with some difficulties with the cylinders with a similar diameter (e.g. diameter 60 mm, 80 mm.) The reason is to be attributed to the small difference in the encoders values, that with the linear mapping, is translated in a small movement, less than 1000 ticks, of the CUFF motors. Regarding the force feedback discrimination task, the subjects were able to recognize precisely the difference between the closed hand and all the other different closure;

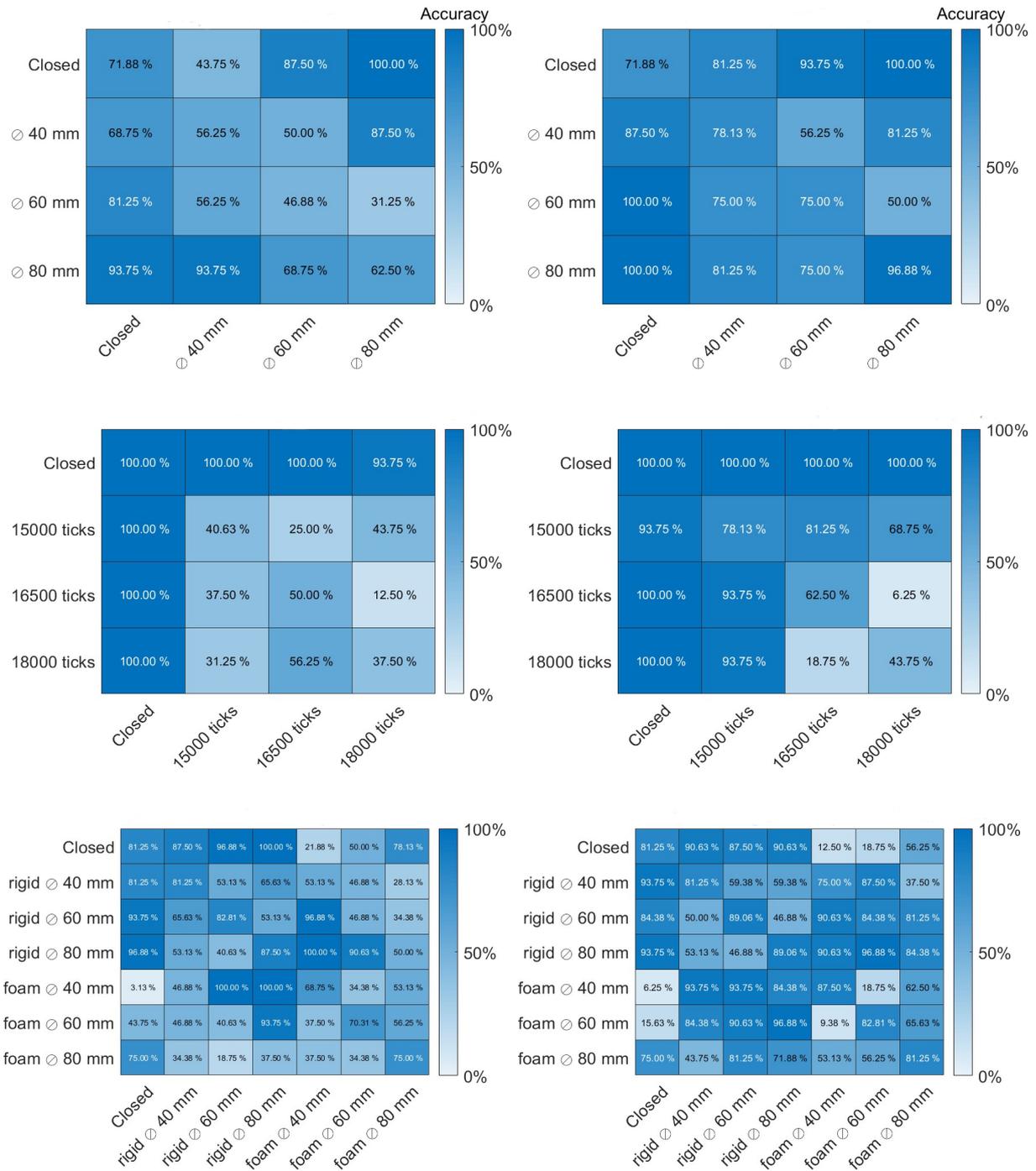

*Figure 11 In these pictures, it is possible to observe the results obtained from the combined stimuli discrimination tasks. First line) results of the linear mapping, second line) results using the logarithmic and exponential mapping, third line) the results of the combined stimuli.*

but there were evident difficulties between all the other pairs of comparison. For the combined stimuli task, as we observed for the proprioception feedback, the participants were able to distinguish the size of the rigid cylinder, and the different size when the comparison was between a rigid and foam cylinder. Overall, we can notice how the percentage of successes is very low in many cases, with a value below 50%. Figure 11(b) and 11(d) (b) showed that the participants of the discrimination task that used the logaritmic and the exponential mapping were able to discriminate the size of all the cylinders, and also to recognize the different force stimuli with a percentage of success near to 100%. We obtained the same positive rate with the combination of the two stimuli. The results confirmed that used a different mapping from the linear one could help to improve the performance and effectiveness of the stimuli delivered by the CUFF's device on the user arm. Looking at the Likert scale results, we can note that in general the participants had no difficulties to wear and use the device, neither in conjunction with the robotic hand (Q1, Q2, Q22 and Q23). About the intuitiveness of the sensory substitution, the stimuli helped discriminate the cylinder size and softness (Q9, Q13, Q18). More specifically, the global impression of the participants was that the stimuli were clear and easy to recognize separately (Q21), but they had some difficulties when receiving the combination of the two stimuli. The subjects, who accomplished the virtual eggs task had the impression of performing better when they receive one of the stimuli by the CUFF with respect to no feedback (Q28, Q29). Regarding Table II, we can note that there are no significant differences between all the four modalities, in terms of the number of fragile objects, moved and broken. What emerged the force stimuli were perceived only in case of break or from the task was that the boxes were too lightweight and over squeeze of the fragile object, instead, the proprioception feedback guided the participants in the first grasp of the object.

# Section VII. Conclusion

We presented the CUFF device, a wearable haptic device able to convey proprioception and force feedback stimuli, and a combination of both. The device was validated with a physical and psychophysical characterization, and tested during an experimental campaign with thirty-two able-bodied participants. Psychometric results showed that the stimuli gen- erated by the device are well perceived by the users wearing it on the upper arm, both in normal force and tangential sliding, with no difference between the two directions of sliding. The experiments focused more on the CUFF, used in conjunction with the SoftHand Pro in size and grasping force discrimination task. The participants were able to use the information coming from the device to distinguish the size, the grasping force and the softness of virtualized cylinders. Survey results showed that it was intuitive and effective, with logarithminc and exponential mapping exhitibing the best results respect to the linear one. Even if the experiments were performed only by able-bodied subjects, the obtained results were encouraging and a more focused study with limb loss participants may be interesting.

## Acknowledgments

This project has received funding from the European Union's Horizon 2020 research and innovation programme under grant agreement No. 688857 (SoftPro). The content of this publication is the sole responsibility of the authors. The European Commission or its services cannot be held responsible for any use that may be made of the information it contains.

The authors would like to thank Nicoletta Colella for her help in the characterization process and Mattia Poggiani for their useful help in the system optimization.